\documentstyle[manuscript,aps]{revtex}

\newcommand{\bib}{\bibitem}
\newcommand{\beq}{\begin{equation}}
\newcommand{\eeq}[1]{\label{#1} \end{equation}}
\newcommand{\beqar}{\begin{eqnarray}}
\newcommand{\eeqar}[1]{\label{#1} \end{eqnarray}}

\newcommand{\bbb}{B^{\prime\prime}}
\newcommand{\aaa}{A^{\prime\prime}}
\newcommand{\fff}{F^{\prime\prime}}
\newcommand{\ddd}{D^{\prime\prime}}
\newcommand{\ccc}{C^{\prime\prime}}
\newcommand{\eee}{E^{\prime\prime}}
\newcommand{\dd}{{\cal D}}

\newcommand{\tr}{{\rm tr}}
\newcommand{\ttt}{t^{\prime}}
\newcommand{\rrr}{r^{\prime}}
\newcommand{\bra}[1]{\langle {#1}|}
\newcommand{\ket}[1]{|{#1}\rangle}
\newcommand{\diag}{{\rm diag}}
\newcommand{\nahm}{\bigtriangleup^{\dagger}\bigtriangleup}
\newcommand{\energy}{\partial_i(\log\det f)}
\newcommand{\hhh}{{\rm H}}
\newcommand{\inter}{\rho_{{\rm int}}}

\preprint{CU-TP-901} 
\title{{\Large \bf Two-Monopole Systems and the
Formation of Non-Abelian Clouds}}
\author{{{\large Changhai Lu} \thanks{chlu@cuphy3.phys.columbia.edu}}\\
\llap{}%
\it{Department of Physics}\\
\it{Columbia University}\\
\it{New York, NY10027, USA}}
\date{June 1998}
\input psfig
\begin{document}
\draft
\maketitle
\begin{abstract}

We study the energy density of two distinct fundamental monopoles 
in $SU(3)$ and $Sp(4)$ theories with an arbitrary mass ratio. Several 
special limits of the general result are checked and verified.
Based on the analytic expression of energy density 
the coefficient of the internal part of the moduli space metric is 
computed, which gives it a nice ``mechanical'' interpretation. 
We then investigate the interaction energy density for both cases. 
By analyzing the contour of the zero interaction energy density we 
propose a detailed picture of what happens 
when one gets close to the massless limit. The study of the interaction
energy density also sheds light on the formation of the 
non-Abelian cloud.

\vspace{5mm}
\noindent PACS number(s): 11.15.-q, 11.27.+d, 14.80.Hv
\end{abstract}
\vfill
\clearpage

\section{Introduction}
Since the pioneering work of t'Hooft and Polyakov \cite{thooft} a 
quarter of a century ago, the study of magnetic monopoles in various 
Yang-Mills-Higgs
theories has become a fruitful direction of research. 
Investigations of these solitonic states have uncovered many deep and 
beautiful structures of gauge theories and greatly improved our 
understanding of those theories.

Magnetic monopoles arise when the Higgs configuration has a non-trivial 
topology at spatial infinity. For a theory with the gauge group $G$ broken
into a residue group $H$, topologically non-trivial configurations are
possible when the second homotopy group of the vacuum manifold, namely
$\Pi_2(G/H)$, is non-trivial. Many works have been performed to understand the 
structure and the metric properties of BPS monopole solutions. It is 
known that when the unbroken gauge group $H$ is Abelian, 
generalization from the original single $SU(2)$ monopole
to multi-monopole systems in arbitrary gauge theories is  
quite straightforward (at least conceptually) \cite{erick}. 

When the unbroken gauge symmetry is non-Abelian, however, the situation becomes
more complicated. Certain fundamental monopoles (namely the monopoles
associated with simple roots of the gauge group) become massless and 
two cases need to be distinguished: the total magnetic
charge carried by monopole is non-Abelian (as the long range magnetic
field transforms non-trivially under unbroken symmetry) or purely Abelian.
In the former case, as was discussed in \cite{abou}, various topological
pathologies appear and prevent us from defining the non-Abelian charge 
globally. On the other hand, when the total magnetic charge is Abelian
(the latter case), there's no topological obstacle, everything behaves
nicely. Therefore the majority of the works 
on monopoles in the presence of non-Abelian unbroken symmetry focus 
this case \cite{dancer} \cite{klee1}\cite{klee2}\cite{erick2}\cite{erick3}.
The modern picture of 
such a case is described by the so called non-Abelian cloud arising from the 
interaction between massless monopoles and massive monopoles. 

In spite of the progress in understanding the field configurations and
the moduli space metrics, the detailed behavior of the interaction that 
accounts for the formation of the non-Abelian cloud is still unclear.
Under the massless limit, a single monopole will spread out and eventually
disappear. This trivial behavior can be significantly changed in the 
presence of massive monopoles. In 
the case where the system carries Abelian charge, the would-be 
massless monopole will lose its identity 
as an isolated soliton once its core region overlaps with
massive monopoles, its size will cease to expand, and its internal 
structure will change in a way that reflects the restored non-Abelian
symmetry. This picture must be distinguished from the 
case when the system carries non-Abelian charge (in the massless limit). 
Having a proper detailed description of the two situations will be helpful
in understanding the formation of the non-Abelian cloud. 
In this paper we will try to address some of these issues. We will
compare the behavior of two monopole systems in $SU(3)$ and $Sp(4)$ 
theories since they are the two simplest models containing the 
interesting contents we are going to study.

The paper is organized as follows: In Sec II, we introduce 
(as the foundation of our calculation) Nahm's 
formalism for the monopole energy density. In Sec III and IV, 
the energy densities
of two (distinct) fundamental monopoles in $SU(3)$ and $Sp(4)$ theories
are calculated and verified in several special cases. In Sec IV
we compute the internal part of the moduli space metric from a 
``mechanical'' point of view. In Sec V we study the formation
of the non-Abelian cloud by analyzing the behavior of the interaction energy 
density. In Sec VI we conclude with some remarks.

\section{Nahm's formalism for the energy density}

As was used in many papers, Nahm's formalism has proved to be a powerful 
tool in calculating many aspects of monopoles. This method is an
analogue of the ADHM construction used in instanton physics \cite{adhm}
and was first proposed by Nahm \cite{nahm}. Recently Nahm's formalism
has been generalized to deal with calorons 
(periodic instantons) \cite{klee3}\cite{baal}, and 
we will use some of the results developed in those works.

Consider the $SU(N)$ Yang-Mills-Higgs system, assuming the 
asymptotic Higgs field along a given direction to be
$\phi_{\infty}={\rm diag}(\mu_1,...,\mu_N)$ 
(with $\mu_1\leq \cdots \leq\mu_N$),
then the Nahm data for the caloron that carries instanton number $k$ are
defined over intervals $(\mu_1,\mu_2)$,...,
$(\mu_{N-1},\mu_N)$, $(\mu_{N},\mu_1+2\pi/\beta)$ (where 
$\beta$ is the circumference along the $S^1$ direction in space 
$R^3\times S^1$) with $\mu_1$ and $\mu_1+2\pi/\beta$ identified. 
In each interval the Nahm data are triplets of $k\times k$ 
Hermitian matrix functions ${\bf T}(t)$ determined by 
Nahm's equations and boundary conditions. In this paper we will
only use the case with $k=1$ for which Nahm data are triplets of constants 
representing the positions of the corresponding constituent
monopoles. It is known that the action density of instantons
(in the usual ADHM method) is given by (it differs from \cite{osb}
by a sign since we choose $F_{\mu\nu}$ to be Hermitian rather than 
anti-Hermitian):
\beq
\rho_s={\rm tr}F^2_{\mu\nu}=\Box\Box\log\det f
\eeq{0}
where $f$ is the inverse operator (whose matrix elements form Green's function)
of $\nahm$ ($\bigtriangleup$ is the 
usual ADHM matrix), and $\Box$ is a four dimensional Laplacian.
For $SU(N)$ calorons 
similar results can be established using the Fourier transformation of the 
original ADHM
method and one has the following formula for the Green's function
$f(t, \ttt)=\bra{t}f\ket{\ttt}$ \cite{klee3}\cite{baal}:
\beq
\left[-\frac{d^2}{dt^2}+|{\bf x}-{\bf T}(t)|^2+
\sum_i |{\bf T}_i-{\bf T}_{i-1}|\delta(t-\mu_i)\right]
f(t, \ttt)=\delta(t-\ttt)
\eeq{2} 
where the sum is taken over all the boundary points between adjacent 
intervals. In order to get the Green's function for monopoles instead 
of calorons, notice that in the constituent monopole picture of calorons, 
an additional type of monopole has been introduced to
neutralize the magnetic charge, so
the usual multi-monopole Green's function can be obtained by moving
the additional monopole to spatial infinity which leads to the following
natural boundary conditions:
\beq
f(\mu_1, \ttt)=f(\mu_N, \ttt)=0
\eeq{2a}
On the other hand, for purely magnetic configurations, the energy density 
is given by (a factor of $1/2$ is omitted for simplicity):
\beq
\rho=\rho_s=\Box\Box\log\det f
\eeq{1}

For later convenience, it would be useful to explore Eq.(\ref{1}) a little 
bit further. Notice that for an operator $f$:
\beq
\log\det f=\tr\log f=-\tr\left[\sum_{n=1}^{\infty}\frac{(1-f)^n}{n}\right]
\eeq{3}
therefore ($i=1, 2, 3$)
\beq
\partial_i\log\det f=\tr\left\{\left[ \sum_{n=0}^{\infty}(1-f)^n\right]
\partial_i f\right\}=\tr(f^{-1}\partial_i f)
\eeq{4}
Further notice that $f=(\nahm)^{-1}$ (Nahm's construction operator
$\bigtriangleup^{\dagger}$ is defined to be
$i\partial_t-i({\bf x}-{\bf T})\cdot 
\sigma$ in each interval), so
\beqar
\tr(f^{-1}\partial_i f) &=& \tr(\nahm\partial_i f)\nonumber\\
&=& \tr [\partial_i(\nahm f)-f\partial_i(\nahm)]\nonumber\\
&=& -\sum_j\left[\int d\ttt f(\ttt, \ttt) \partial_i|{\bf x}
-{\bf T}_j|^2\right]
\eeqar{5}
So finally we have obtained a convenient formula for the energy density:
\beq
\partial_i\log\det f=-\sum_j\left[ 
\partial_i|{\bf x}-{\bf T}_j|^2
\int d\ttt f(\ttt, \ttt) \right]
\eeq{green}
where $\partial_i|{\bf x}-{\bf T}_j|^2$ 
has been moved out from the integration since in each interval
${\bf T}_j$ does not depend on $\ttt$ (for the 
$k=1$ case only).
Equation (\ref{green}) together with Eqs.(\ref{2}) (\ref{2a}) (\ref{1}) 
(in Eq.(\ref{1}) use three dimensional Laplacian $\bigtriangleup$ 
to replace $\Box$) forms a framework to compute energy density of 
monopole systems considered in this paper.

\section{Two fundamental monopoles in $SU(3)$ theory}

The root diagram of $SU(3)$ theory is shown in Fig 1:

\begin{figure}
\hspace{1.9in}
\psfig{file=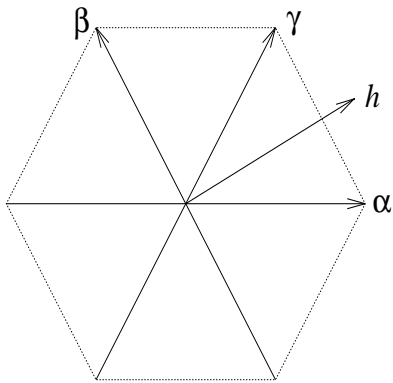,height=2in,width=2in}
\caption{Root diagram of $SU(3)$ theory}
\end{figure}

We will consider the system formed by one $\alpha$ and one $\beta$
monopole, $h$ in the graph refers
to the Higgs direction along which $\beta$ monopole is massless. 

\subsection{Energy density}

We choose $\phi_{\infty}$ (along a given direction) to be 
$\diag (-1-\mu, 2\mu, 1-\mu)$ (with $-1/3\leq \mu\leq 1/3$), so the 
masses of the two fundamental monopoles are;
\beq
m_{\alpha}=1+3\mu ; \hspace{5mm} m_{\beta}=1-3\mu
\eeq{6}
Without losing generality, we can place the $\alpha$-monopole on the
origin and $\beta$-monopole on $(0, 0, \dd)$ which is equivalent to 
choosing ${\bf T}(t)=(0, 0, 0)$ for $t\in (-1-\mu, 2\mu)$
and ${\bf T}(t)=(0, 0, \dd)$ for $t\in (2\mu, 1-\mu)$.
Applying Eqs.(\ref{2}) (\ref{2a}) to this case we have 
\beq
\left[ -\frac{d^2}{dt^2}+|{\bf x}-{\bf T}(t)|^2+\dd\delta (t-2\mu) \right] 
f(t, \ttt)=\delta (t-\ttt)
\eeq{7a}
\beq
f(-1-\mu, \ttt)=f(1-\mu, \ttt)=0
\eeq{7b}
It's easy to see from these equations
that the Green's function has the following form:
\begin{itemize}
\item Case A: $-1-\mu <\ttt <2\mu$
\beqar
f(t, \ttt)=\left\{\begin{array}{lc}
A\sinh[r(t+1+\mu)] & (-1-\mu < t <\ttt)\\
B\sinh(rt)+C\cosh(rt) & (\ttt < t <2\mu)\\
D\sinh[\rrr (1-\mu -t)] & (2\mu < t <1-\mu)\end{array}\right.
\eeqar{8a}
\item Case B: $2\mu < \ttt <1-\mu$
\beq
f(t, \ttt)=\left\{\begin{array}{lc}
A^{\prime}\sinh[r(t+1+\mu)] & (-1-\mu < t < 2\mu)\\
B^{\prime}\sinh(\rrr t)+C^{\prime}\cosh(\rrr t) & (2\mu < t <\ttt)\\
D^{\prime}\sinh[\rrr (1-\mu -t)] & (\ttt < t < 1-\mu)\end{array}\right.
\eeq{8b}
\end{itemize}
where $r=\sqrt{x_1^2+x_2^2+x_3^2}$, $\rrr =\sqrt{x_1^2+x_2^2+
(x_3-\dd)^2}$ are distances from two monopoles and the coefficients
$A, B, C, A^{\prime}, B^{\prime}, C^{\prime}$ are all functions
of $\ttt$. 

In each case Eq.(\ref{7a}) also implies the usual boundary conditions 
(which we won't bother writing down) concerning 
the continuity of $f(t, \ttt)$ and the jumps of $\partial_t f(t, \ttt)$
at each point where the argument of the $\delta$-functions becomes zero. 
All the coefficients can be computed from those boundary conditions. 
It's helpful to notice that Eq.(\ref{green}) makes use of $f(t, \ttt)$ only 
in the form of $\int d\ttt f(\ttt, \ttt)$ which is equal to
\beqar
\int d\ttt f(\ttt, \ttt)&=&\int^{2\mu}_{-1-\mu}d\ttt A(\ttt)
\sinh [r(\ttt+1+\mu)] \nonumber \\
&+&\int^{1-\mu}_{2\mu}d\ttt D^{\prime}(\ttt)
\sinh[\rrr (1-\mu -\ttt)]
\eeqar{9}
and so we only need $A$ and $D^{\prime}$. Computing them using boundary
conditions and putting
into Eqs.(\ref{9}) and (\ref{green}) one obtains the following result:
\beqar
\partial_i (\log\det f)&=& -\hat{r}_i\frac{rp\sinh p\sinh q+A_1(p\cosh p 
-\sinh p)}{rM}\nonumber\\
&-&\hat{r}^{\prime}_i\frac{\rrr q\sinh q\sinh p+A_2(q\cosh q 
-\sinh q)}{\rrr M}
\eeqar{10}
where $p, q, A_1, A_2, M$ are defined as:
\beq
p=m_{\alpha}r;\hspace{5mm} q=m_{\beta}\rrr
\eeq{11}
\beq
A_1=\dd\sinh q+\rrr\cosh q;\hspace{5mm} A_2=\dd\sinh p
+r\cosh p
\eeq{12}
\beq
M=\dd\sinh p\sinh q+r\cosh p\sinh q+\rrr\sinh p \cosh q
\eeq{13}
From Eq.(\ref{10}) one can also derive the regularized determinant of Green's
function to be
\beq
(\det f)_{{\rm reg}}=\frac{r\rrr}{M}
\eeq{12a}
which is defined in the sense that it is finite and gives the same
$\partial_i (\log\det f)$ and energy density $\rho$ through:
\beq
\partial_i (\log\det f)=\partial_i [\log (\det f)_{{\rm reg}}]
\eeq{12b}
\beq
\rho=\bigtriangleup\bigtriangleup \log (\det f)_{{\rm reg}}
\eeq{12c}
Three typical configurations are 
shown in Fig 2 (we plot it on the $x-z$ plane since the configurations
are axially symmetric). 

\begin{figure}
\hspace{1in}\vspace{-1in}\psfig{file=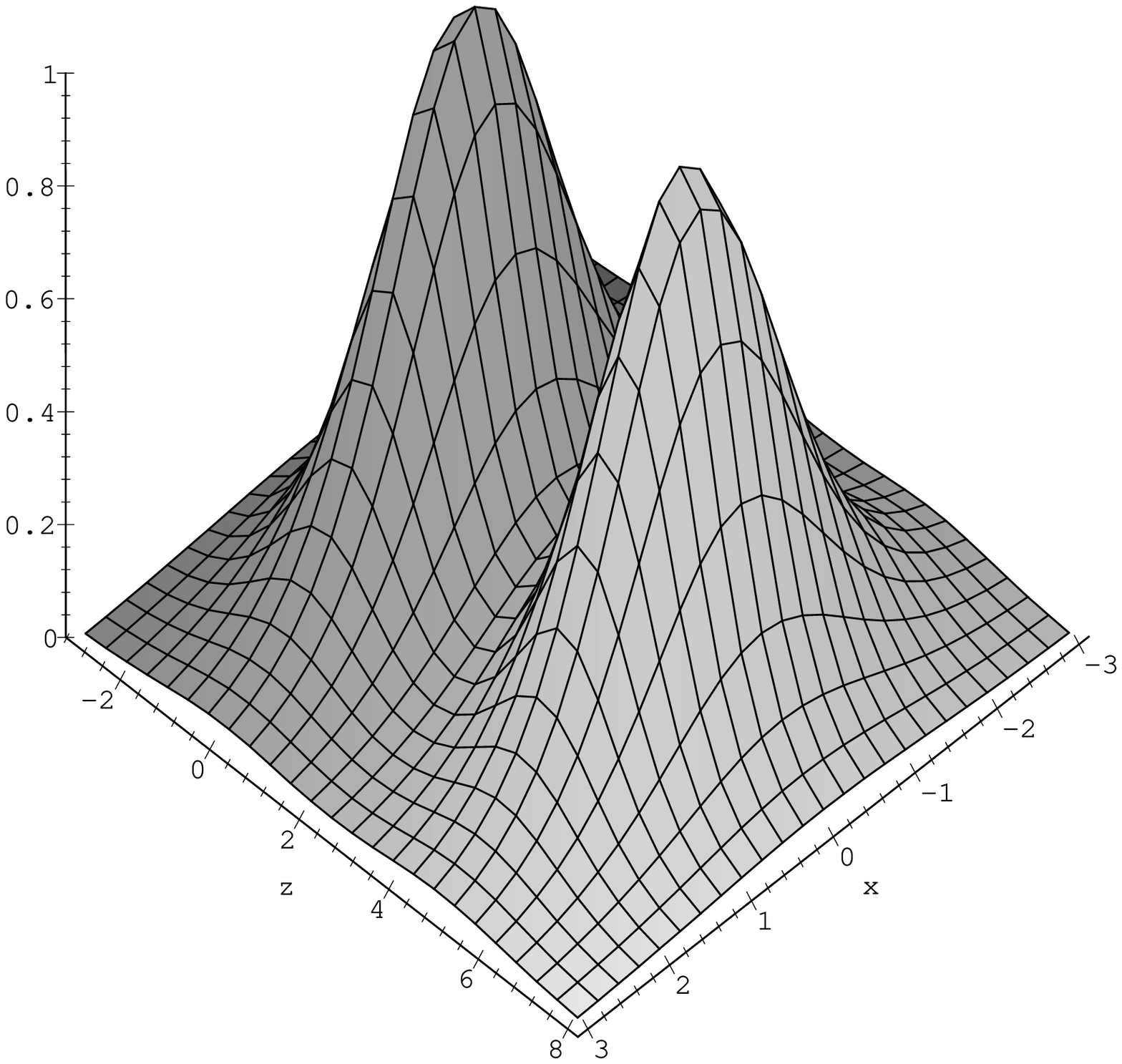,width=4in,height=3in}

\hspace{1in}\vspace{-1in}\psfig{file=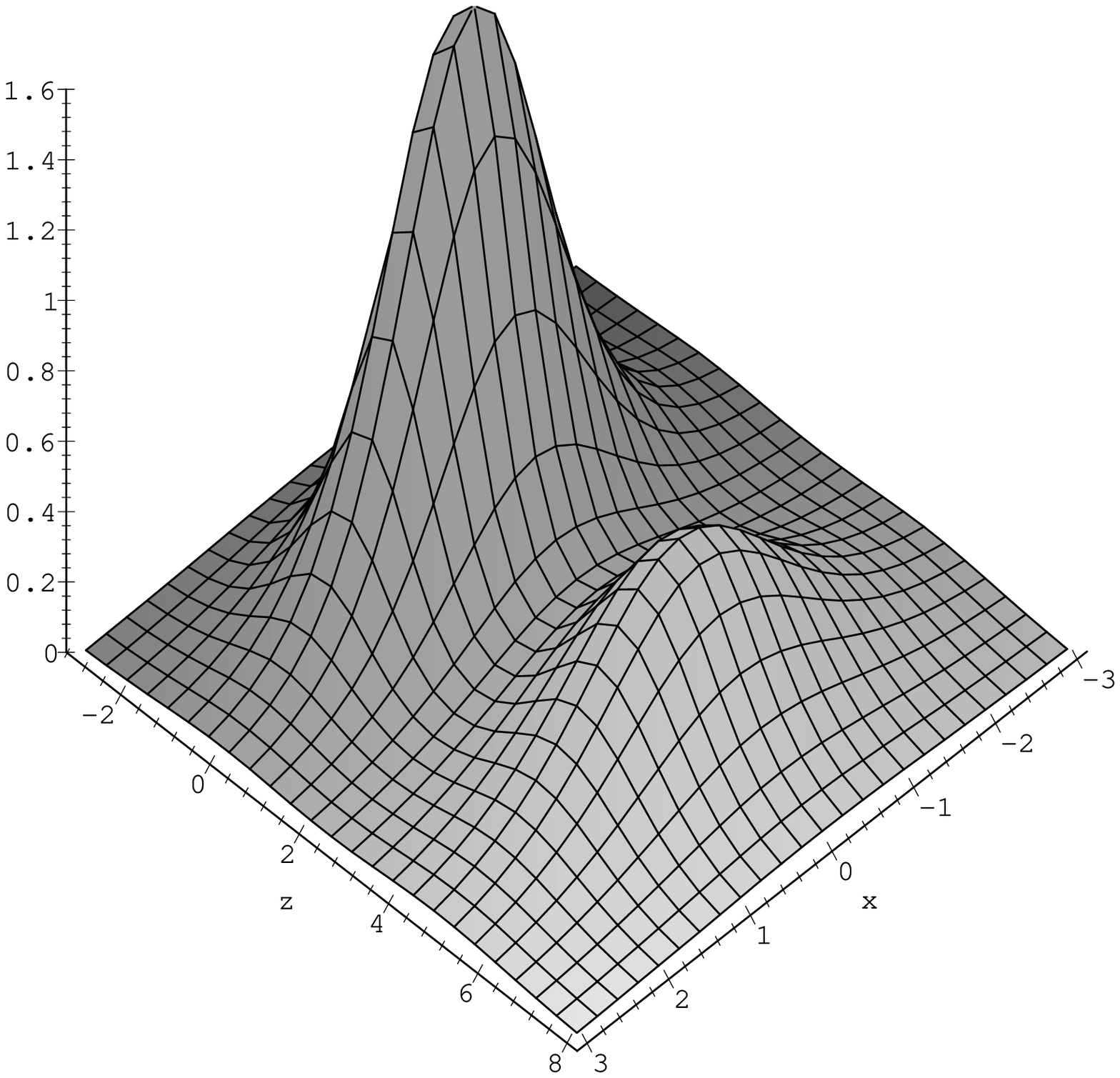,width=4in,height=3in}

\hspace{1in}\vspace{-1in}\psfig{file=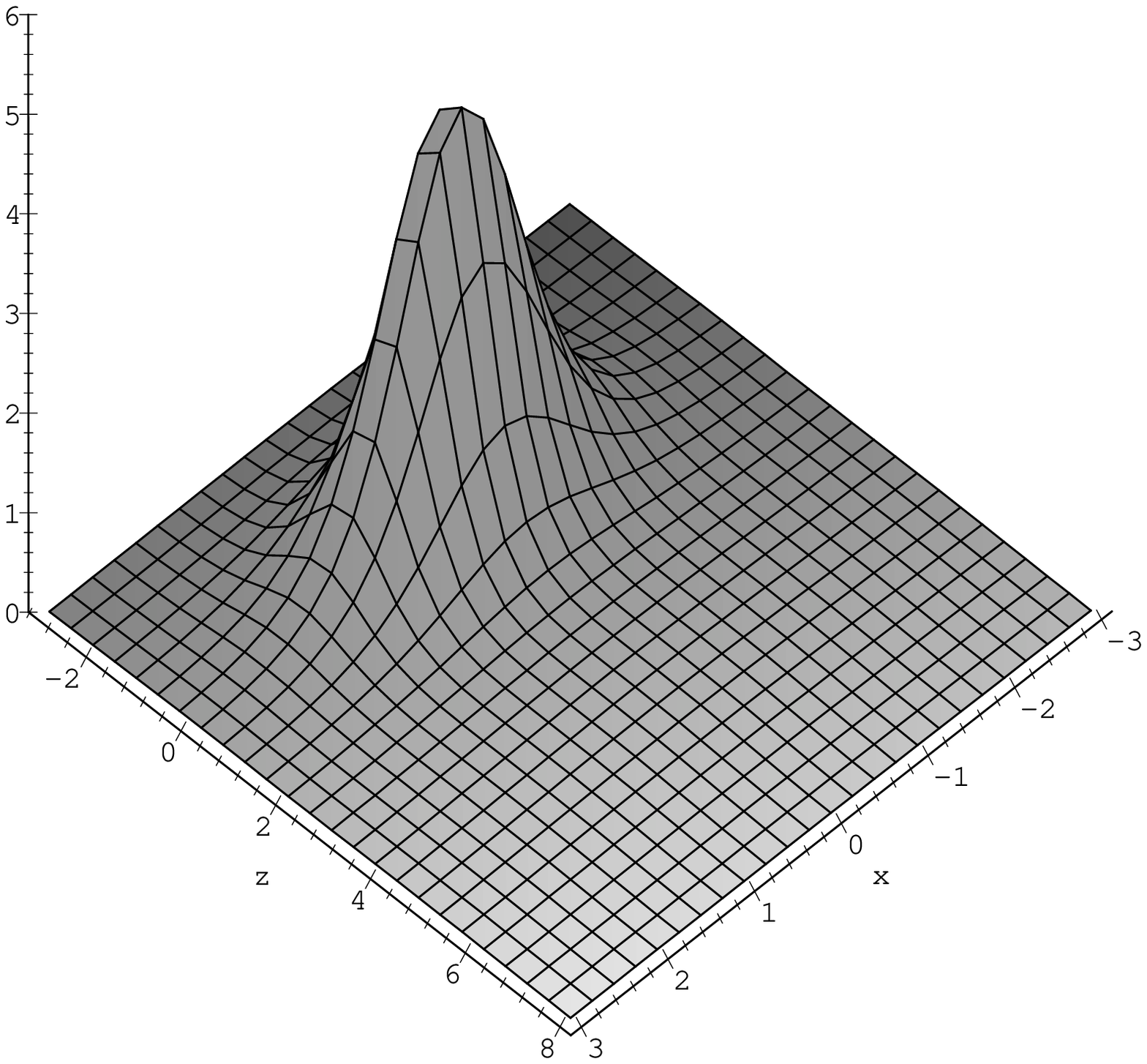,width=4in,height=3in}

\vspace{1in}
\caption{Energy density of two fundamental monopoles 
on the $x-z$ plane in $SU(3)$ theory.
$\alpha$ and $\beta$ monopoles are located at $(0, 0)$ and $(0, 5)$. The 
mass ratio $m_{\alpha}/m_{\beta}$ (from top to bottom) is chosen to be
1, 1.35 and 3}
\end{figure}

\subsection{Various limits of the energy density}
In this subsection let's check certain limits of the general form
of the energy density, this serves as a partial verification of the result
obtained in last subsection. 
\begin{enumerate}
\item $\dd=0$ case (two monopoles are on top of each other):

In this case we expect (and will see) that 
the resulting energy density is the same as that of an
$SU(2)$-embedded monopole with mass $m=m_{\alpha}+m_{\beta}=2$.
Since two monopoles are overlapped, so $r=\rrr$ and one has:
\beq
M=r\sinh (2r)
\eeq{14}
So (using Eqs.(\ref{12a})(\ref{12b})) 
\beq
\energy=-\hat{r}_i\hhh(2r)
\eeq{15}
where $\hhh(2r)$ is the $m=2$ ($m$ is the mass parameter) case of the 
single monopole function defined as
\beq
\hhh(mr)=m\left[\coth(mr)-\frac{1}{mr}\right]
\eeq{16}
From Eq.(\ref{15}), one can further get:
\beq
\bigtriangleup (\log\det f)=\partial_i\energy=\hhh^2(2r)-4
\eeq{17}
and therefore
\beq
\rho=\bigtriangleup\bigtriangleup (\log\det f)=\bigtriangleup [\hhh^2(2r)]
\eeq{17a}
which is fully compatible (in the suitable convention of normalization)
with another well known formula \cite{ward}:
\beq
\rho=\bigtriangleup (\tr \phi^2)
\eeq{18}
since $\phi\propto \hhh(2r)$ for single $SU(2)$-embedded monopole
with mass $m=2$.

\item Massless limit:

This is the case when one of the monopoles becomes massless (we will 
investigate this limit in more detail in Sec IV). In our convention
this happens when $\mu=\pm 1/3$. Without losing generality, we choose
$\mu=1/3$ (so the $\beta$-monopole is massless), then $p=2r$, $q=0$ and
\beq
M=\rrr\sinh (2r)
\eeq{19}
and therefore
\beq
\energy=-\hat{r}_i\hhh(2r)
\eeq{20}
which (as the $\dd=0$ case) again leads to the energy density of 
a single $SU(2)$ monopole with mass $m=2$. Notice $m_{\alpha}=2$ 
in this case, so such a result means that the 
$\beta$-monopole doesn't affect the energy density
in the massless limit (which, as we will see,
is very different from $Sp(4)$ case).

\item Removing one monopole:

Again, without losing generality, let's move the $\beta$-monopole away so
that $\rrr\rightarrow\infty$, the dominant term in $M$ is
\beq
M\sim\rrr\exp(m_{\beta}\rrr)\sinh(m_{\alpha}r)
\eeq{21}
which leads to
\beq
\energy=-\hat{r}_i\hhh(m_{\alpha}r)-m_{\beta}\hat{r}^{\prime}_i
\eeq{22}
Since $\hat{r}^{\prime}_i$ represents a constant vector at this limit, so
Eq.(\ref{22})
is exactly what one expects for a single monopole with mass $m_{\alpha}$.
\end{enumerate}

\section{Two fundamental monopoles in $Sp(4)$ theory}

$Sp(4)$ (or equivalently $SO(5)$) theory is the simpest theory 
to study the non-Abelian cloud. The root diagram is shown in Fig.3:
\begin{figure}[h]
\hspace{1.9in}
\psfig{file=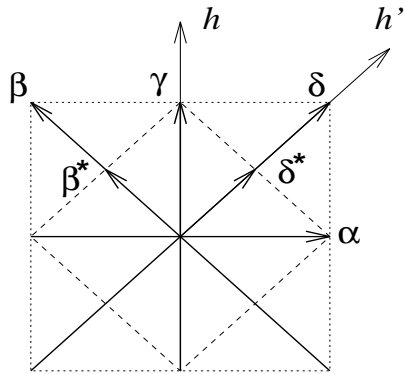,height=2in,width=2in}
\caption{Root diagram of $Sp(4)$ theory. $\beta^*$,
$\delta^*$ are co-roots}
\end{figure}
Two cases containing massless monopoles have been studied before.
When the asymptotic Higgs field is along the $h$ direction, the Abelian 
configuration is made of two massive $\beta$-monopoles and one massless
$\alpha$-monopole \cite{klee2}, when the asymptotic Higgs field is along the
$h^{\prime}$ direction, the Abelian configuration is made of one massive
$\alpha$-monopole and one massless $\beta$-monopole \cite{klee1}. 
In this section we will consider the general energy density for a one
$\alpha$ one $\beta$ monopole system (with arbitrary mass ratio) which
will help us to see how the massless limit would be eventually 
achieved.

\subsection{Energy density}

We choose $\phi_{\infty}$ (along a given direction) to be 
$\diag(-1, -\mu, \mu, 1)$ (with $0\leq\mu\leq 1$), so the masses of
the fundamental monopoles are:
\beq
m_{\alpha}=2-2\mu;\hspace{5mm} m_{\beta}=2\mu
\eeq{23}
The monopole locations are chosen to be the same as in the $SU(3)$ case. As we 
know, in Nahm's method $Sp(4)$ theory is embedded into $SU(4)$ theory  
whose Nahm data satisfy symmetric constraints \cite{klee2}, in our 
convention the $SU(4)$ Nahm data 
are given by ${\bf T}(t)=(0, 0, 0)$ for 
$t\in (-1, -\mu)\cup (\mu, 1)$ and ${\bf T}(t)=(0, 0, \dd)$
for $t\in (-\mu, \mu)$. Applying Eqs.(\ref{2}) (\ref{2a}) to this case we have:
\beq
\left[ -\frac{d^2}{dt^2}+|{\bf x}-{\bf T}(t)|^2+\dd\delta (t+\mu) 
+\dd\delta (t-\mu) \right] f(t, \ttt)=\delta (t-\ttt)
\eeq{24}
\beq
f(-1, \ttt)=f(1, \ttt)=0
\eeq{25}
The Green's function satisfying these equations has the following form:
\begin{itemize}
\item Case A: $-1<\ttt <-\mu$
\beqar
f(t, \ttt)=\left\{\begin{array}{lc}
A\sinh [r(t+1)] & (-1 <t<\ttt)\\
B\sinh(rt)+C\cosh(rt) & (\ttt <t<-\mu)\\
D\sinh(\rrr t)+E\cosh(\rrr t) & (-\mu <t<\mu)\\
F\sinh[r(1-t)] & (\mu <t<1)\end{array}\right.
\eeqar{26a}
\item Case B: $-\mu <\ttt <\mu$
\beqar
f(t, \ttt)=\left\{\begin{array}{lc}
A^{\prime}\sinh[r(t+1)] & (-1<t<-\mu)\\
B^{\prime}\sinh(\rrr t)+C^{\prime}\cosh(\rrr t) & (-\mu <t<\ttt)\\
D^{\prime}\sinh(\rrr t)+E^{\prime}\cosh(\rrr t) & (\ttt <t<\mu)\\
F^{\prime}\sinh[r(1-t)] & (\mu <t<1)\end{array}\right.
\eeqar{26b}
\item Case C: $\mu <\ttt <1$
\beqar
f(t, \ttt)=\left\{\begin{array}{lc}
\aaa\sinh[r(t+1)] & (-1<t<-\mu)\\
\bbb\sinh(\rrr t)+\ccc\cosh(\rrr t) & (-\mu <t<\mu)\\
\ddd\sinh(r t)+\eee\cosh(r t) & (\mu <t<\ttt)\\
\fff\sinh[r(1-t)] & (\ttt <t<1)\end{array}\right.
\eeqar{26c}
\end{itemize}
where $r$ and $\rrr$ have the same meaning as before, and 
$A,...,F^{\prime\prime}$ are all functions of $\ttt$. 

Similarly as in the $SU(3)$ case, we only need some of the coefficients ($A$, 
$B^{\prime}$, $C^{\prime}$, $\ddd$) which (as well as all other coefficients)
can be determined by boundary conditions coming from the 
$\delta$-functions in Eqs.(\ref{24}). Compute them and putting them into
Eqs.(\ref{green}) one obtains:
\beq
\energy = -\hat{r}_i \frac{PA_1+QA_2}{\cosh r P+\sinh r Q}
-\hat{r}^{\prime}_i\left( \frac{MB_1}{L}+\frac{LB_2}{M}\right)
\eeq{27}
where the following functions are introduced:
\beq
u=(1-\mu)r;\hspace{5mm} v=\mu r;\hspace{5mm} w=\mu\rrr
\eeq{28a}
\beqar
A_1&=&\frac{-\cosh r +\cosh(u-v)}{r}+2(1-\mu)\sinh r\label{28aa}\\
A_2&=&\frac{-\sinh r -\sinh(u-v)}{r}+2(1-\mu)\cosh r
\eeqar{28b}
\beq
B_1=\mu+\frac{\sinh(2\mu\rrr)}{2\rrr}; \hspace{5mm}
B_2=\mu-\frac{\sinh(2\mu\rrr)}{2\rrr}
\eeq{28c}
\beqar
L&=&\dd\sinh u\cosh w+r\cosh u\cosh w+\rrr\sinh u\sinh w\\
M&=&\dd\sinh u\sinh w+r\cosh u\sinh w+\rrr\sinh u\cosh w\\
N_1&=&\dd\sinh v\sinh w-r\cosh v\sinh w+\rrr\sinh v\cosh w\\
N_2&=&\dd\cosh v\sinh w-r\sinh v\sinh w+\rrr\cosh v\cosh w\\
N_3&=&\dd\sinh u\sinh(2w)+r\cosh u\sinh(2w)+\rrr\sinh u\cosh(2w)
\eeqar{28d}
\beq
P=\rrr\sinh v M-N_1N_3;\hspace{5mm} Q=-\rrr\cosh v M+N_2N_3
\eeq{28e}
The reguralized determinant of The Green's function from Eq.(\ref{27}) is
\beq
(\det f)_{{\rm reg}}=\frac{r^2\rrr}{LM}
\eeq{28f}
Three typical configurations are shown in Fig 4.

\begin{figure}

\hspace{1in}\vspace{-1in}\psfig{file=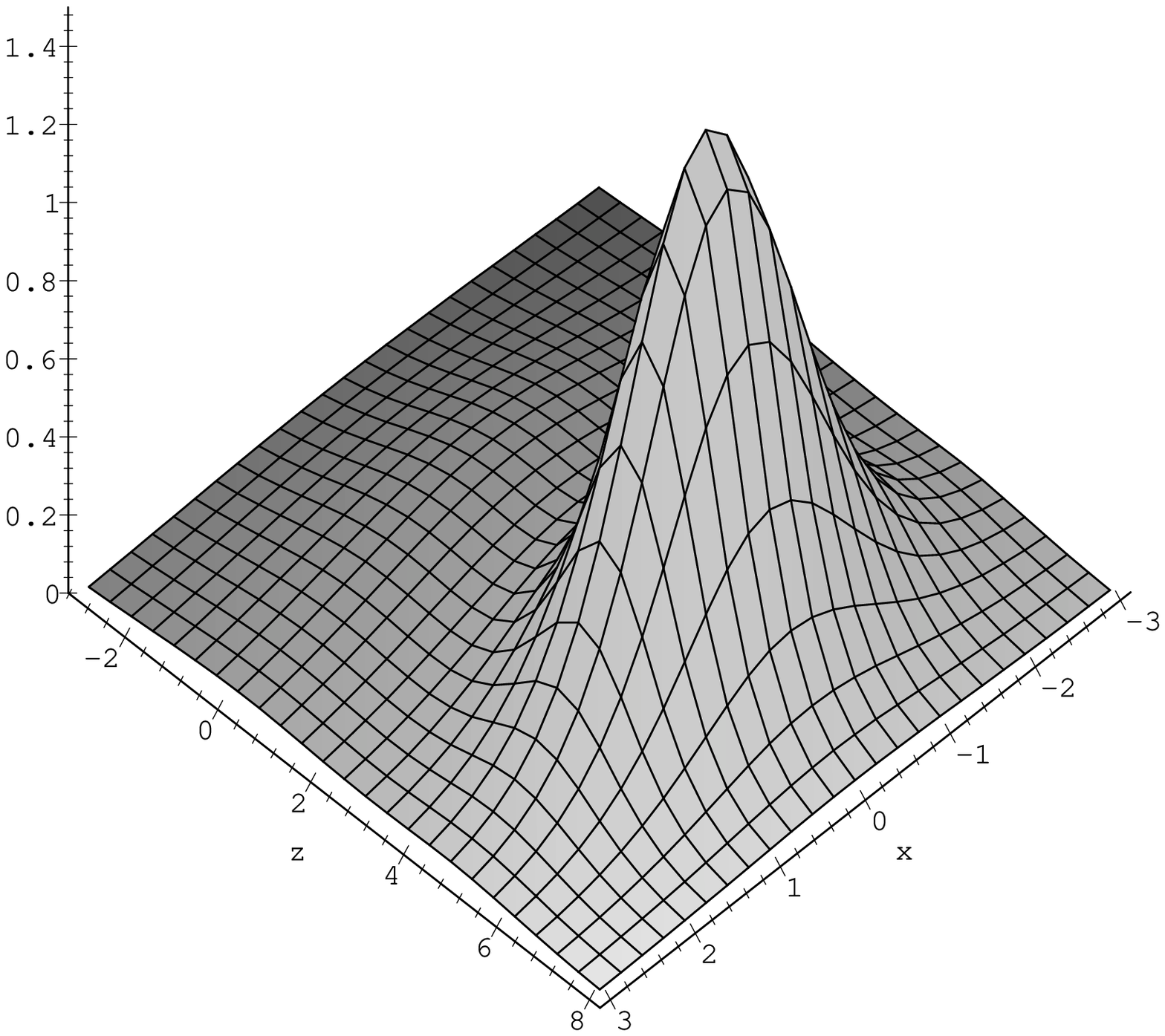,width=4in,height=3in}

\hspace{1in}\vspace{-1in}\psfig{file=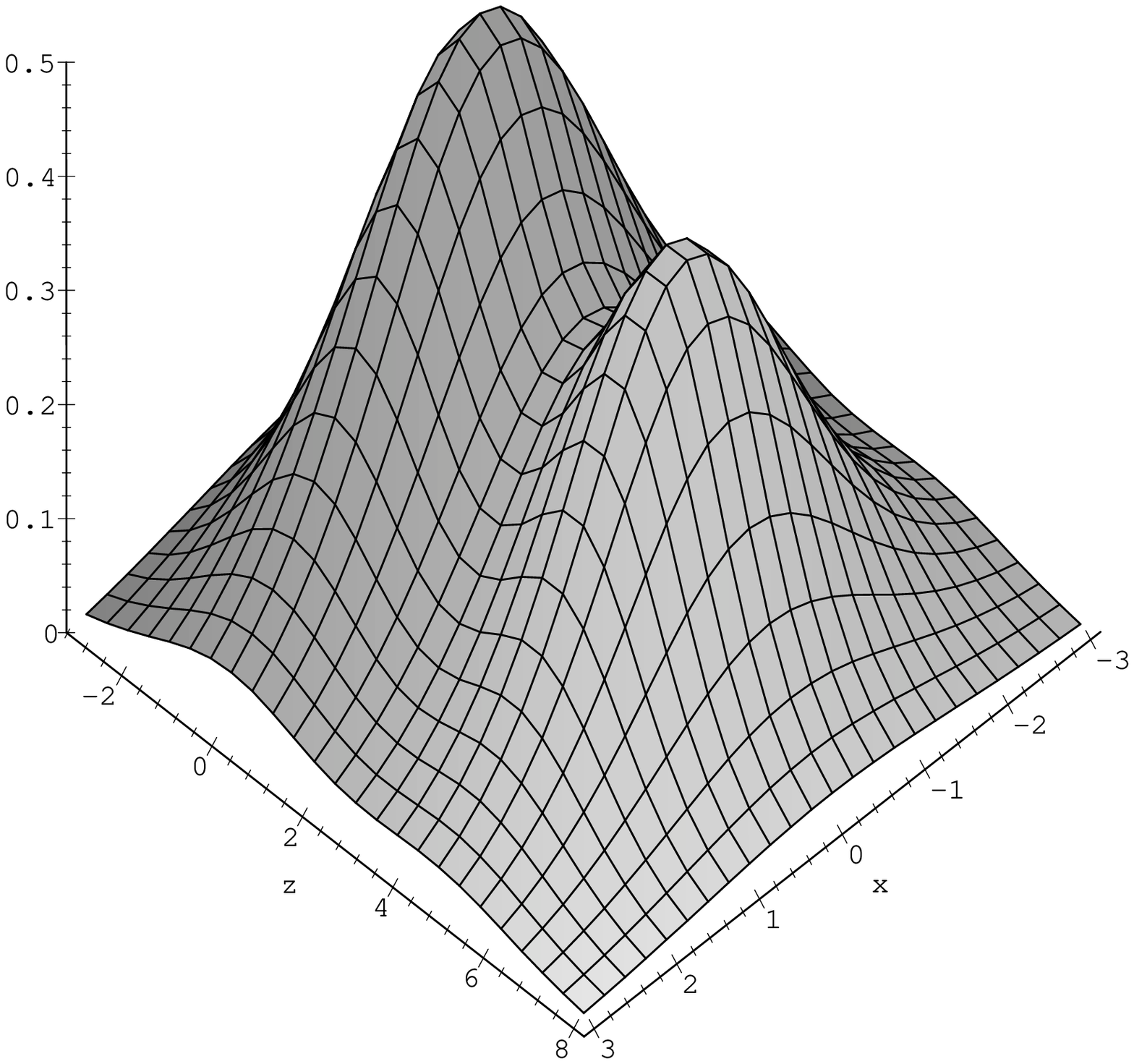,width=4in,height=3in}

\hspace{1in}\vspace{-1in}\psfig{file=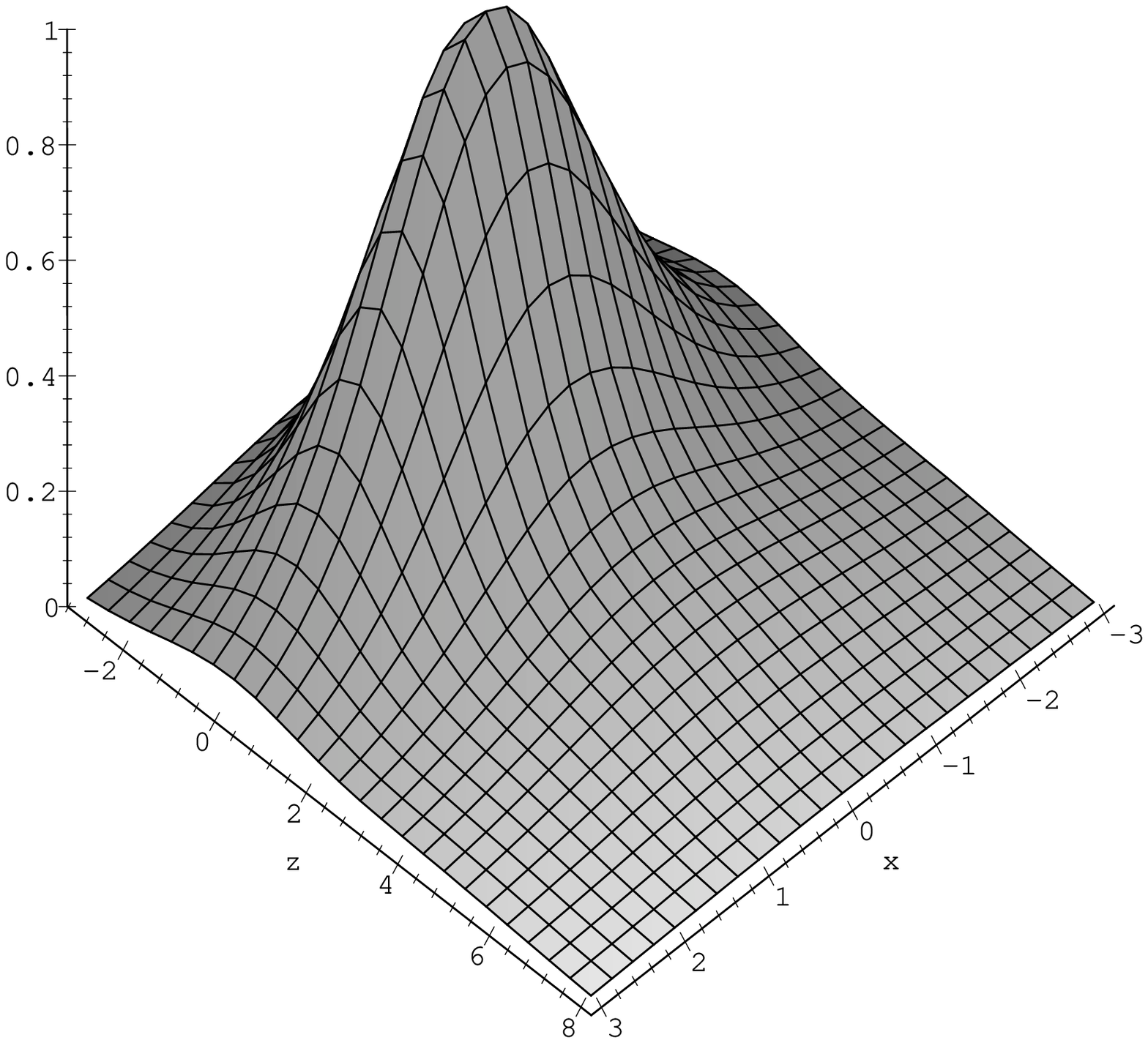,width=4in,height=3in}

\vspace{1in}

\caption{Energy density of two fundamental monopoles 
on the $x-z$ plane in $Sp(4)$
theory. $\alpha$ and $\beta$ monopoles are located at 
$(0, 0)$ and $(0, 5)$. Mass ratio $m_{\alpha}/m_{\beta}$ 
(from top to bottom) is chosen to be 1, 2 and 4}

\end{figure}

One can see that the energy density of the two fundamental
monopoles in $Sp(4)$ theory is not symmetric, this is because here
$\alpha$ and $\beta$ monopoles have different energy distributions.
As we know, the mass of $SU(2)$-embedded $\gamma$-monopoles in arbitrary
gauge theory is given by $h\cdot \gamma^*$ while
the scale of such monopoles is determined by
$1/(h\cdot\gamma)$ ($\gamma$ 
is the root to which the $SU(2)$-embedded monopoles are associated,
$h$ is the asymptotic Higgs direction). 
As a consequence, in $Sp(4)$ theory when $\alpha$ and $\beta$ monopoles
have the same mass, the scale of the $\beta$-monopole is only half of the 
scale of the $\alpha$-monopole therefore the maximal energy density 
is eight times that of the $\alpha$-monopole, this is why in the plotting
with $m_{\alpha}:m_{\beta}=1$ one can hardly see the $\alpha$-monopole.

\subsection{Various limits of the energy density}

The expression of the energy density of the
two fundamental monopoles in $Sp(4)$ 
theory is much more complicated than in the $SU(3)$ case, to see its 
correctness, let's check several special cases.
\begin{enumerate}
\item $\dd=0$ case (two monopoles are on top of each other):

In this case we again expect an $SU(2)$-embedded monopole with mass
$m=m_{\alpha}+m_{\beta}=2$. Since $r=\rrr$, one gets
\beq
L=r\cosh r;\hspace{5mm} M=r\sinh r 
\eeq{29a}
which generates the familar result:
\beq
\energy=-\hat{r}_i\hhh(2r)
\eeq{30}
which is the same as Eq.(\ref{15}), so in this limit the result is 
exactly what we expected.

\item Massless limit:

We are interested in the limit when the total magnetic charge is 
Abelian which is realized by $\mu\rightarrow 0$ (when the $\beta$-monopole 
becomes massless). In this limit one has
\beq
L=\dd\sinh r+r\cosh r;\hspace{5mm} M=\rrr\sinh r
\eeq{31a}
which leads to 
\beq
\energy=-\hat{r}_i\frac{2r^2\cosh(2r)+2\dd r\sinh(2r)-r\sinh(2r)
-2\dd\cosh(2r)+2\dd}{r^2\sinh(2r)+\dd r\cosh(2r)-\dd r}
\eeq{32}
On the other hand, the Higgs configuration in this limit is given in 
\cite{erick2}\cite{erick3}, 
in that work, the Higgs configuration of $N-1$ fundamental
monopoles is calculated in the symmetry breaking pattern 
$SU(N)\rightarrow U(1)\times SU(N-2)\times U(1)$. The relevant $Sp(4)$
Higgs configuration can be obtained by taking the $N=4$ case and putting
two massive monopoles on top of each other. In a proper normalization
the result of \cite{erick2} (after simplification) can be written as
(where $\sigma_r=\sigma\cdot\hat{r}$)
\beq
\phi=\frac{1}{\sqrt{2}}\left(\begin{array}{cc}
\hhh(2r)\sigma_r & \sqrt{\frac{2\dd\tanh^2 r}{\sinh(2r)(r+\dd\tanh r)}}\\
\sqrt{\frac{2\dd\tanh^2 r}{\sinh(2r)(r+\dd\tanh r)}} &
\frac{2\dd r-\dd\sinh(2r)}{2r\cosh^2r(r+\dd\tanh r)}\sigma_r 
\end{array}\right)
\eeq{32a}
which leads to 
\beq
\tr\phi^2=\hhh^2(2r)+\frac{4\dd\tanh^2r}
{\sinh(2r)(r+\dd\tanh r)}+\left[\frac{2\dd r-\dd\sinh(2r)}
{2r\cosh^2r(r+\dd\tanh r)}\right]^2
\eeq{33}
As we did before, in order to show that Eqs.(\ref{1}) 
(\ref{32}) and (\ref{18}) (\ref{33}) give the 
same result it is sufficient to show that the difference between
$\partial_i\energy$ and $\tr\phi^2$ is constant. This 
is messy, but one can verify that
\beq
\partial_i\energy=\tr\phi^2-4
\eeq{34}
so the massless limit works out correctly. Unlike the $SU(3)$ case, the 
massless limit of $Sp(4)$ theory is rather non-trivial (the non-Abelian
cloud coming into playing). 

\hspace{5mm} One can easily show that when the 
$\alpha$-monopole becomes massless
(in that case the total magnetic charge is non-Abelian), the 
situation is similar to the $SU(3)$ case, namely the energy density is 
equal to the energy density of a single $\beta$-monopole.

\item Removing the $\alpha$-monopole:

Now let's check what happens when one monopole is removed.
Since the two monopoles are not symmetric in the $Sp(4)$ case so we check
them separately. When $\dd\rightarrow\infty$ but keeping $\rrr$ finite
(therefore $r\rightarrow\infty$), the $\alpha$-monopole is removed.
The leading contributions of $L, M$ are:
\beq
L\sim r\exp[(1-\mu)r]\cosh(\mu\rrr);\hspace{5mm}
M\sim r\exp[(1-\mu)r]\sinh(\mu\rrr)
\eeq{35}
Putting these into Eqs.(\ref{28f}) and (\ref{12b}) one obtains:
\beq
\energy=-\hat{r}^{\prime}_i\hhh(2\mu\rrr)-m_{\alpha}\hat{r}_i
\eeq{36}
This (notice that $\hat{r}_i$ is a constant vector at this limit)
represents an $SU(2)$-embedded monopole with mass $m=m_{\beta}$ 
which is what we are expecting.

\item Removing the $\beta$-monopole:

When $\dd\rightarrow\infty$,
$\rrr\rightarrow\infty$, the $\beta$ monopole is removed.
The leading contributions of $L, M$ are
\beqar
L\sim M\sim\rrr\exp(\mu\rrr)\sinh [(1-\mu)r]
\eeqar{37}
which lead to (again, the second term is a constant vector at this limit) 
\beq
\energy=-2\hat{r}\hhh [(1-\mu)r]-m_{\beta}\hat{r}^{\prime}_i
\eeq{38}
The energy density coming from this expression is the same as two 
directly superposed $SU(2)$ monopoles with 
mass $m=1-\mu=m_{\alpha}/2$. This is consistent
with the $SU(4)$ picture introduced in 
the discussion of the massless limit, namely the 
$\alpha$ monopole can be considered as the two overlapped 
$SU(4)$ monopoles whose energy densities are simply added since
they are non-interacting. This also gives a direct
demonstration of our discussion at the end of subsection A.
\end{enumerate}

\subsection{From the moment of inertia to the moduli space metric}

Since we have the analytic form of energy density, we are now able
to compute the internal part of the moduli space metric using 
a nice ``mechanical'' interpretation. The idea of using a mechanical 
interpretation can be traced to Manton's original 
work \cite{manton} (where the concept of the moduli space metric 
itself was introduced
by comparing the action of the monopole system and mechanical system) and 
was used in certain arguments
in recent works \cite{klee2}\cite{irwin}. In this
subsection we will consider the massless limit of our $Sp(4)$ system. 
The metric in this case is known to be \cite{klee1} (changed into our
convention):
\beq
ds^2=md{\bf x}^2+\frac{16\pi^2}{m}d\chi^2+\frac{4\pi}{\dd} d\dd^2
+4\pi\dd (\sigma_1^2+\sigma_2^2+\sigma_3^2)
\eeq{40}
where $m$ is the total mass (which is just the mass of the $\alpha$-monopole
in this case) of the system, $\sigma_1, \sigma_2$ and $\sigma_3$ are
1-forms defined as
\beqar
\sigma_1&=&-\sin\psi d\theta +\cos\psi\sin\theta d\phi\\
\sigma_2&=&\cos\psi d\theta +\sin\psi\sin\theta d\phi\\
\sigma_3&=&d\psi +\cos\theta d\phi
\eeqar{41}
with the Euler angles $\theta, \phi$ and $\psi$ having periodicities
$\pi, 2\pi$ and $4\pi$, respectively. It's interesting to notice
that the last term in Eq.(\ref{40}) has a ``mechanical interpretation'' as
the rotational energy associated with the massless cloud with the coefficient
$4\pi\dd$ playing the role of the moment of inertia of the cloud. 
To see this let's calculate the moment of inertia of the non-Abelian cloud.
As we know, for a spherically symmetric system the moment of inertia
tensor has the form $I_{ij}=I\delta_{ij}$ with
\beq
I=\frac{2}{3}\int dV r^2\rho
\eeq{42}
Since the rotation of the 
cloud is actually the gauge rotation in internal space, 
we should remove the gauge invariant part $\rho(\dd=0)$ (this
is an $SU(2)$-embedded $\gamma$-monopole which is gauge invariant), so
the effective energy density relevant to the internal rotation is $\rho(\dd)-
\rho(0)$. Using the result obtained in the last subsection one can
verify that 
\beqar
I&=&\frac{2}{3}\int dVr^2[\rho(\dd)-\rho(0)]\nonumber\\
&=&\frac{8\pi}{3}\int dr r^4[\rho(\dd)-\rho(0)]\nonumber\\
&=&16\pi\dd
\eeqar{43}
which leads to a term $16\pi\dd (d\omega_1^2+d\omega_2^2+d\omega_3^2)$
in the moduli space metric. Here 1-forms $d\omega_1, d\omega_2$ and 
$d\omega_3$ are defined in the group space of $SU(2)$ (namely $S^3$). 
To compare with the last term of Eq.(\ref{40}) one notices that 
\beq
\sigma_1^2+\sigma_2^2+\sigma_3^2=d\theta^2+d\phi^2+
d\psi^2+2\cos\theta d\phi d\psi
\eeq{44}
therefore the volume of $(\theta, \phi, \psi)$ space is ($g=\det(g_{ij})$
is the determinant of metric matrix coming from Eq.(\ref{44})) 
\beq
{\cal V}=\int \sqrt{|g|}d\theta d\phi d\psi=16\pi^2
\eeq{45}
Since the volume of 
group space $S^3$ is $2\pi^2$ the two sets of 1-forms are related
by $\sigma_i=2d\omega_i$ 
($i=1, 2, 3)$. Taking this into account we see that 
$16\pi\dd (d\omega_1^2+d\omega_2^2+d\omega_3^2)$
computed from the moment of inertia is in accordance with the last term  
in Eq.(\ref{40}) obtained using other methods. 

\section{Interaction energy density and the formation of the non-Abelian
cloud}

In previous sections we have calculated the energy density of two
monopole systems in $SU(3)$ and $Sp(4)$ theory. We've already known
that when one approaches the massless limit, the situations are very
different depending on the total magnetic charge. If the total magnetic
charge is non-Abelian, the resulting energy density is simply the same
as the energy density of the massive monopole. When the total
magnetic charge is purely Abelian, however, the energy density 
distribution is deeply affected by the existence of massless monopoles.
In the latter case, there is a non-Abelian cloud surrounding the massive
monopole, neutralizing the non-Abelian components of the magnetic charge. In 
this section we want to have a close look at the evolution of the energy 
density when one approaches the massless limit.

There's no unique choice of quantity to describe the formation of the 
non-Abelian cloud, nor is there any unambiguous definition of the non-Abelian 
cloud itself. But physically there is no doubt that it is the interaction
between massive and massless monopoles that determines the 
behavior of the system, including the formation of the cloud.
Our strategy is to study the interaction energy density defined as
\beq
\rho_{{\rm int}}=\rho_{{\rm total}}-\rho_{\alpha}-\rho_{\beta}
\eeq{50}
where $\rho_{\alpha}$ and $\rho_{\beta}$ are the energy density of 
isolated $\alpha$ and $\beta$ monopoles. 
$\inter$ describes the change of energy distribution caused by the 
interaction between two monopoles. In particular we will look at 
the contour of the zero interaction energy density which gives
information on where interaction gathers energy and from where 
it extracts energy. 

We have used Maple to generate numerical data and plotted several 
typical contours (shown in Figs 5, 6)
for $SU(3)$ and $Sp(4)$ theories (the region enclosed by the contour
has a positive interaction energy density).

\begin{figure}[h]
\hspace{1.5in}\psfig{file=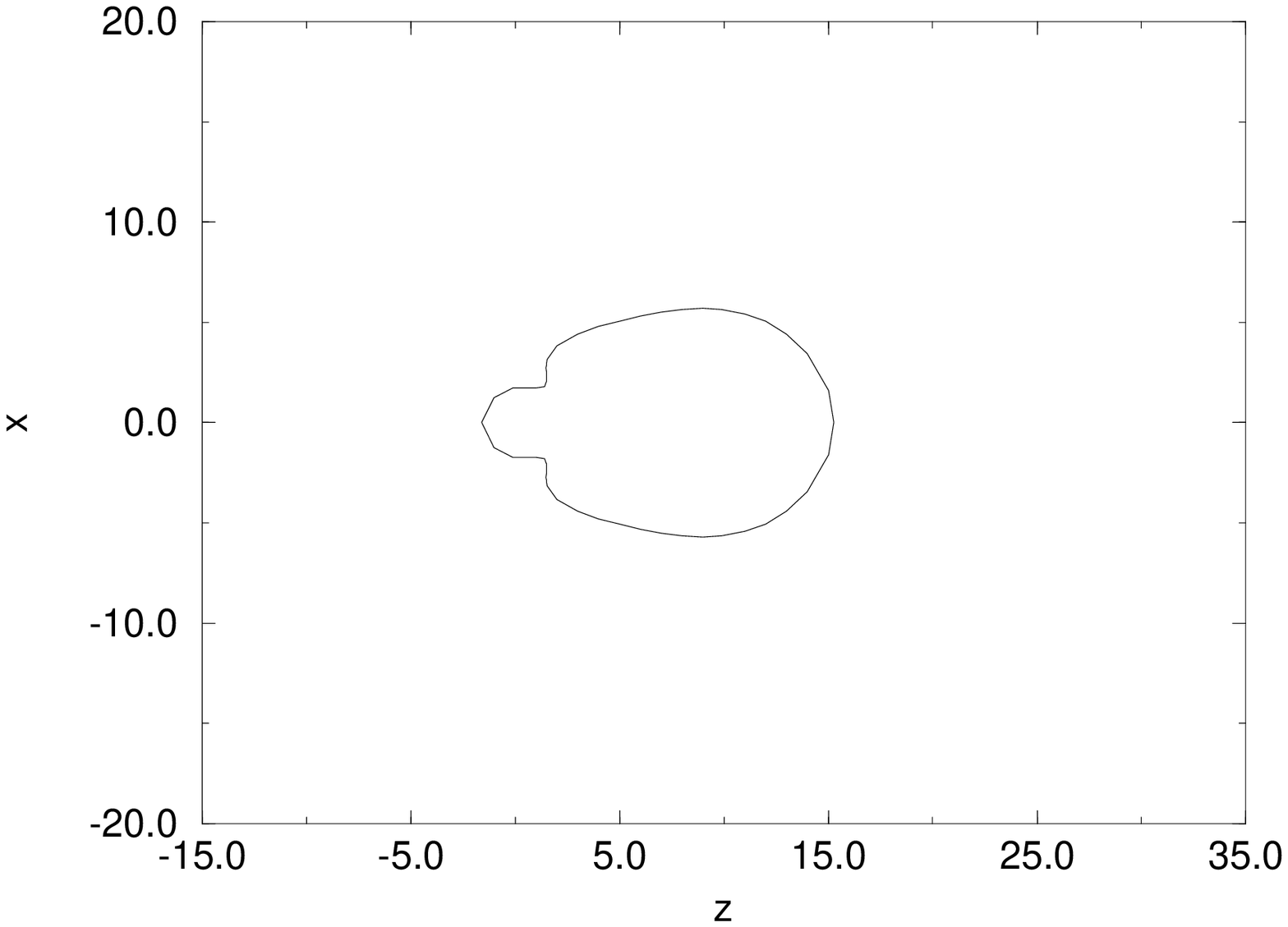,width=3in,height=2.5in}

\hspace{1.5in}\psfig{file=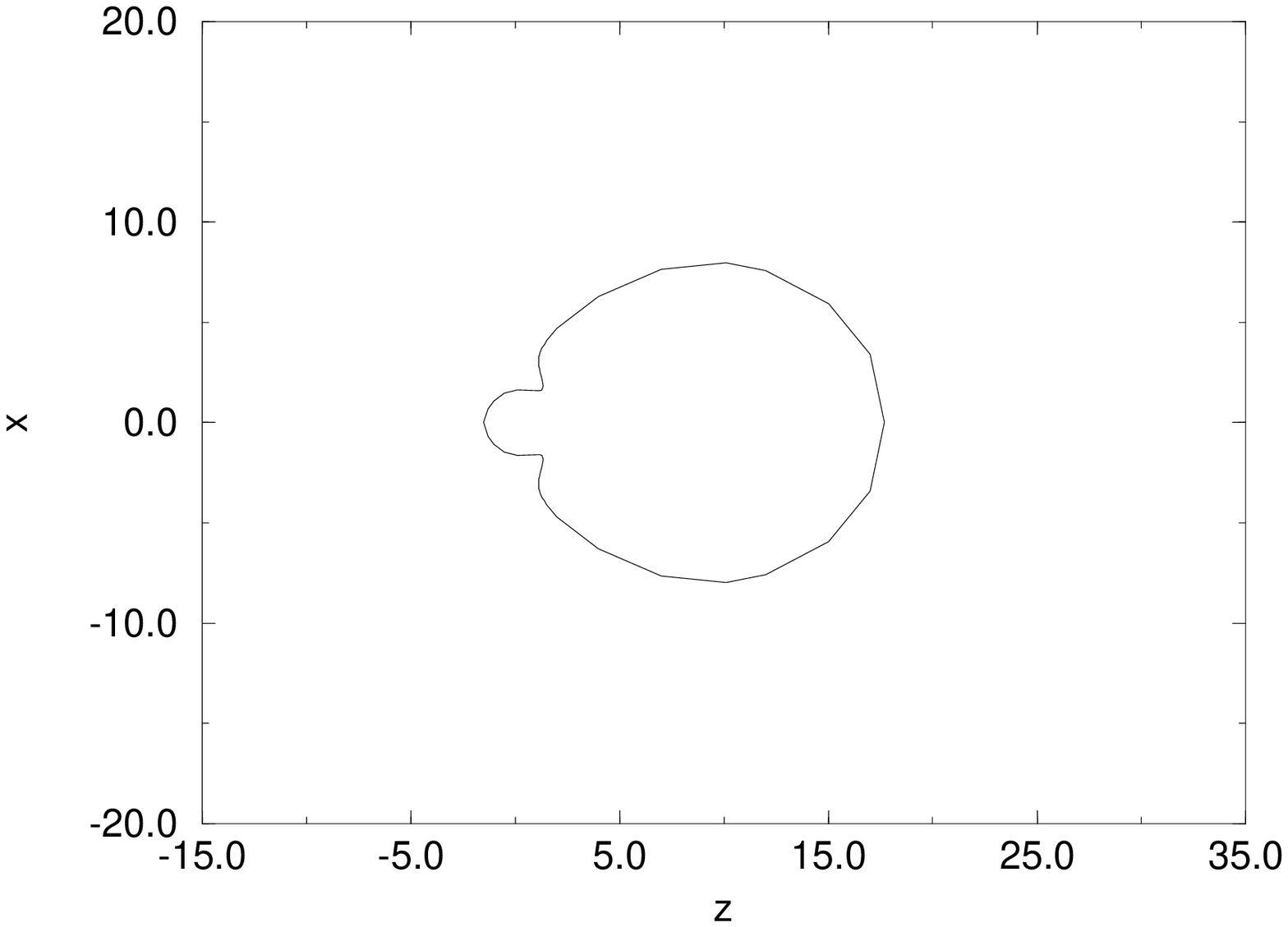,width=3in,height=2.5in}

\hspace{1.5in}\psfig{file=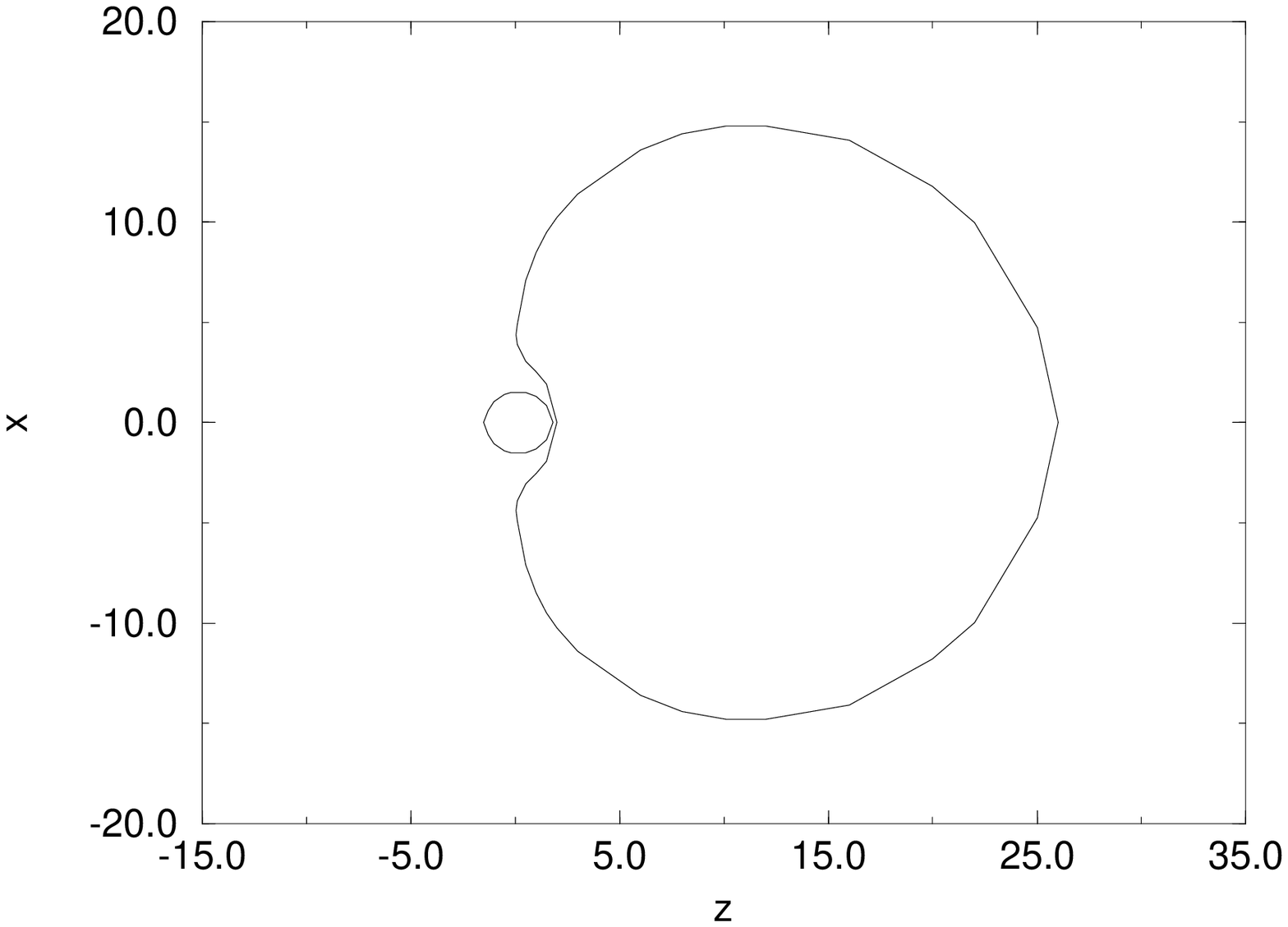,width=3in,height=2.5in}

\hspace{1.5in}\psfig{file=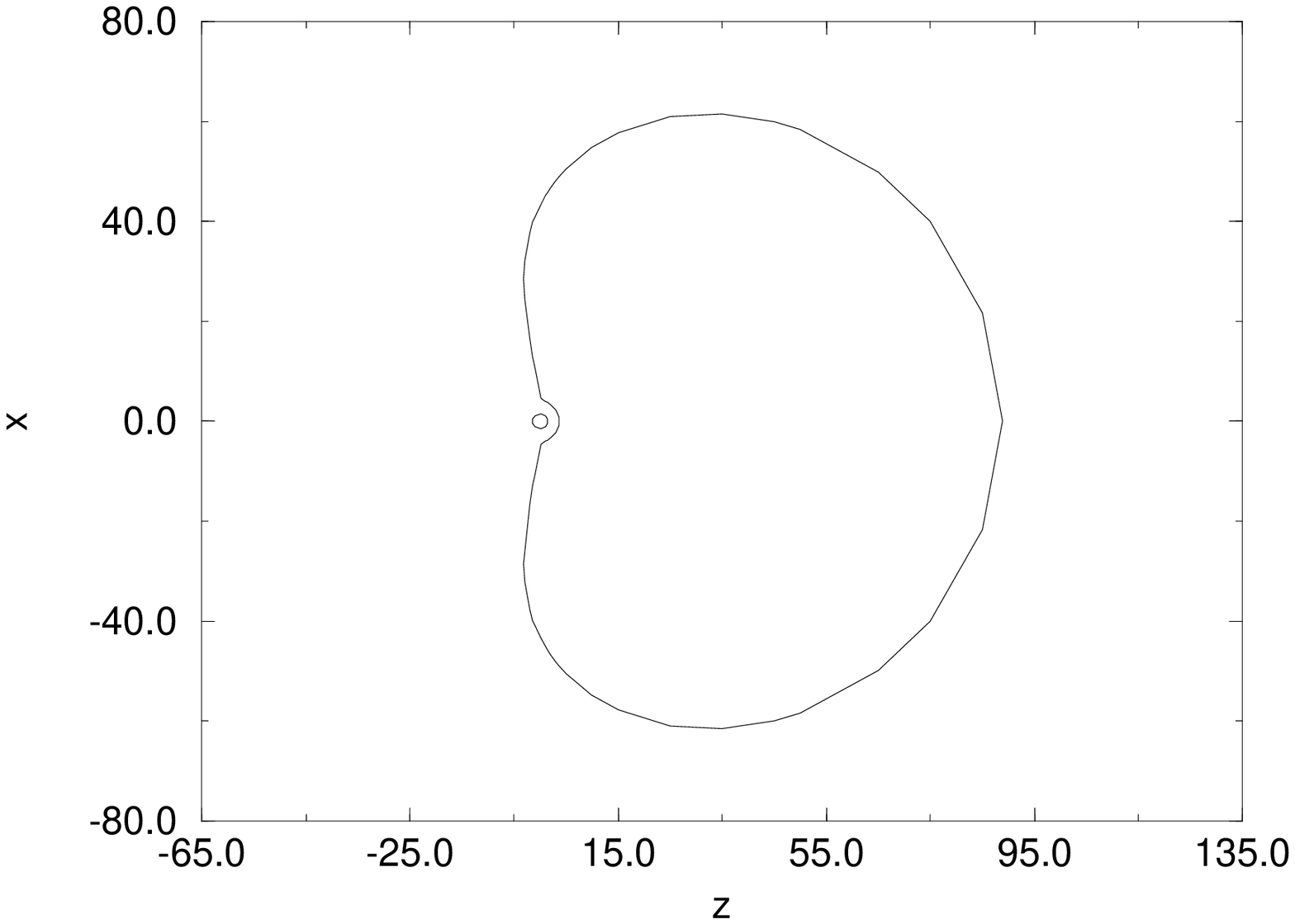,width=3in,height=2.5in}

\caption{Contour diagrams of the zero interaction energy density 
on the $x-z$ plane for $SU(3)$
theory. $\alpha$ and $\beta$ monopoles are located at $(0, 0)$ and
$(0, 10)$, mass ratios $m_{\alpha}:m_{\beta}$ are choosen to be
(from top to bottom) 4, 7, 19, 199}
\end{figure}

\begin{figure}[h]
\hspace{1.5in}\psfig{file=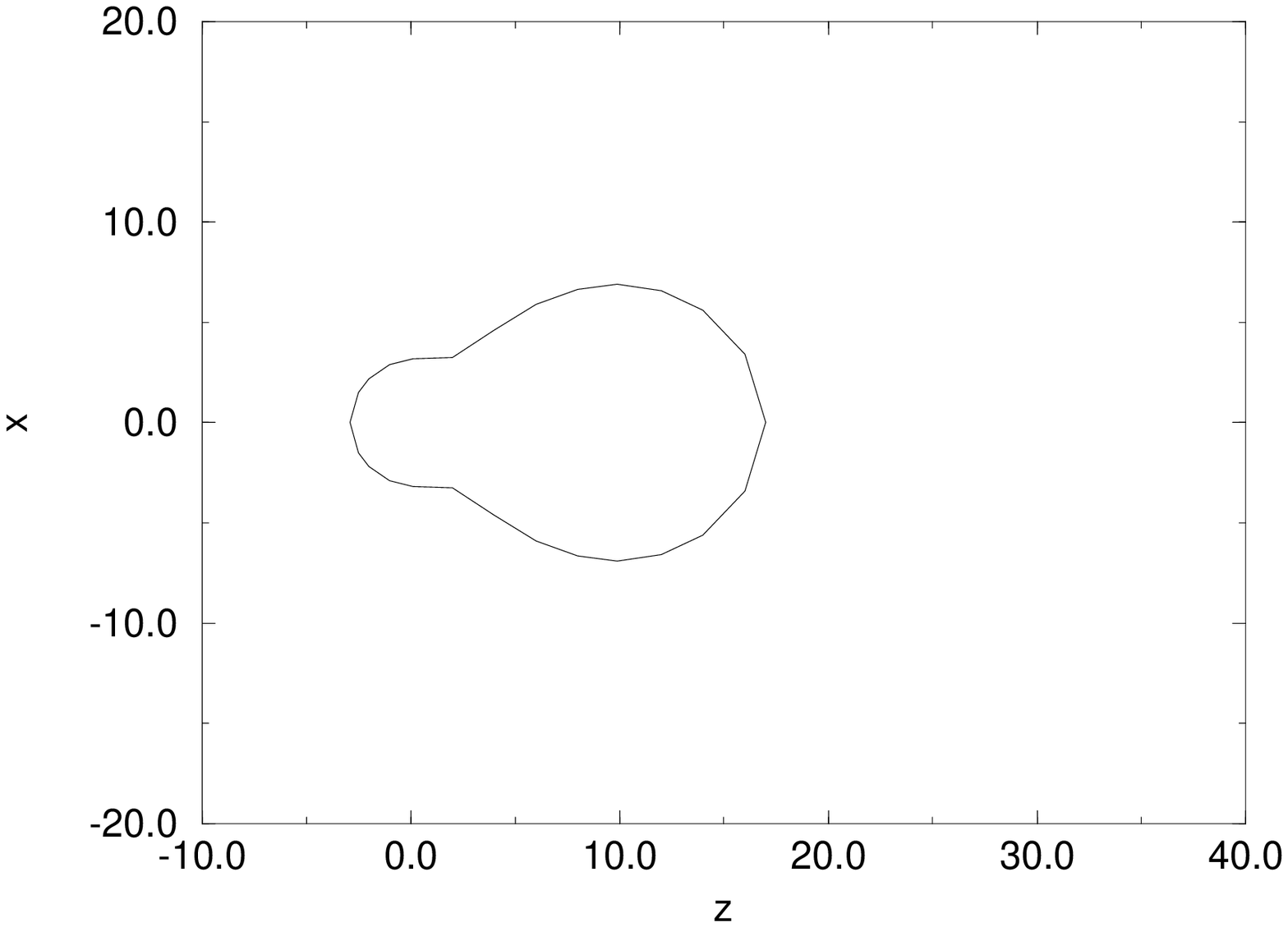,width=3in,height=2.5in}

\hspace{1.5in}\psfig{file=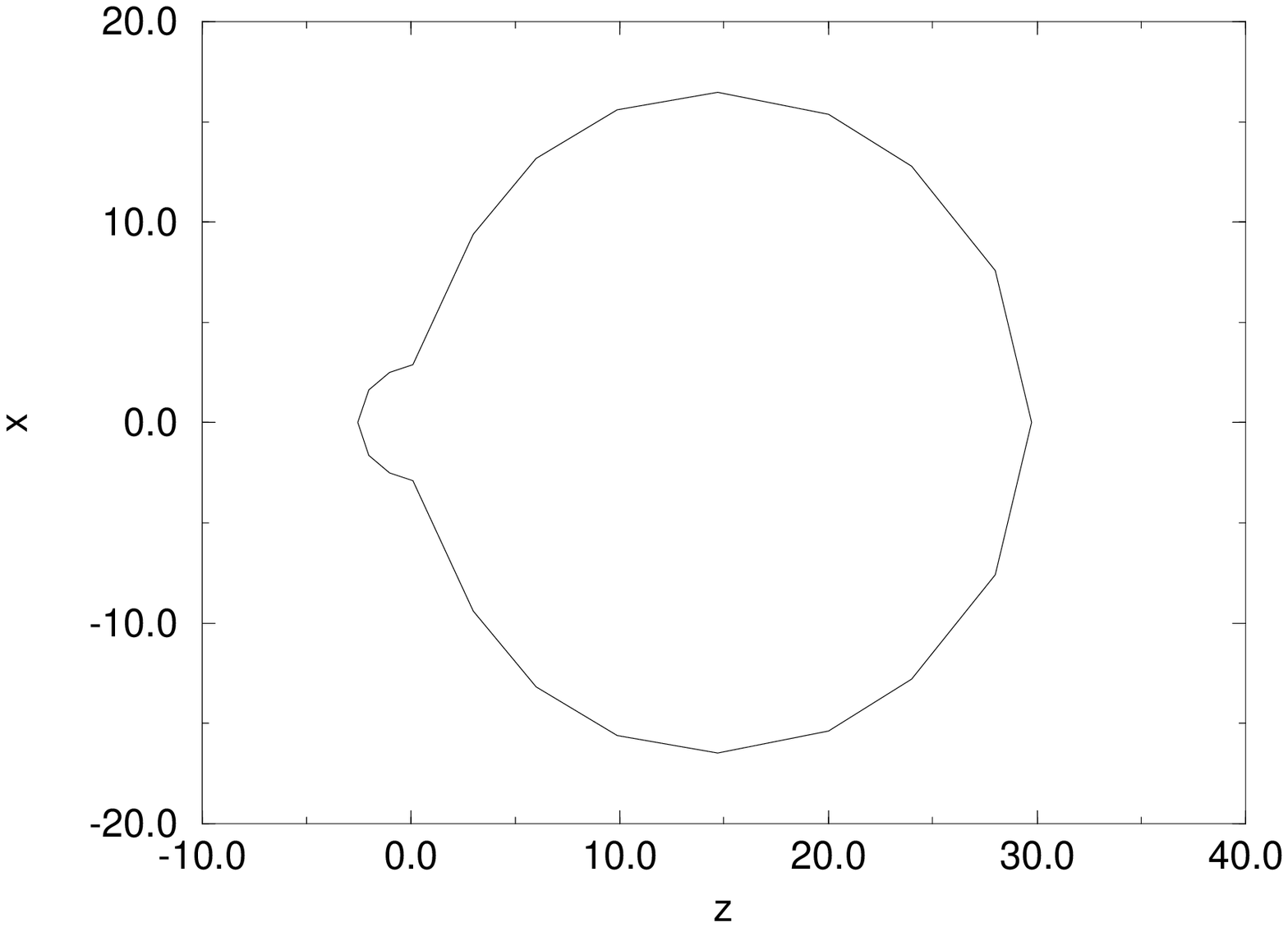,width=3in,height=2.5in}

\hspace{1.5in}\psfig{file=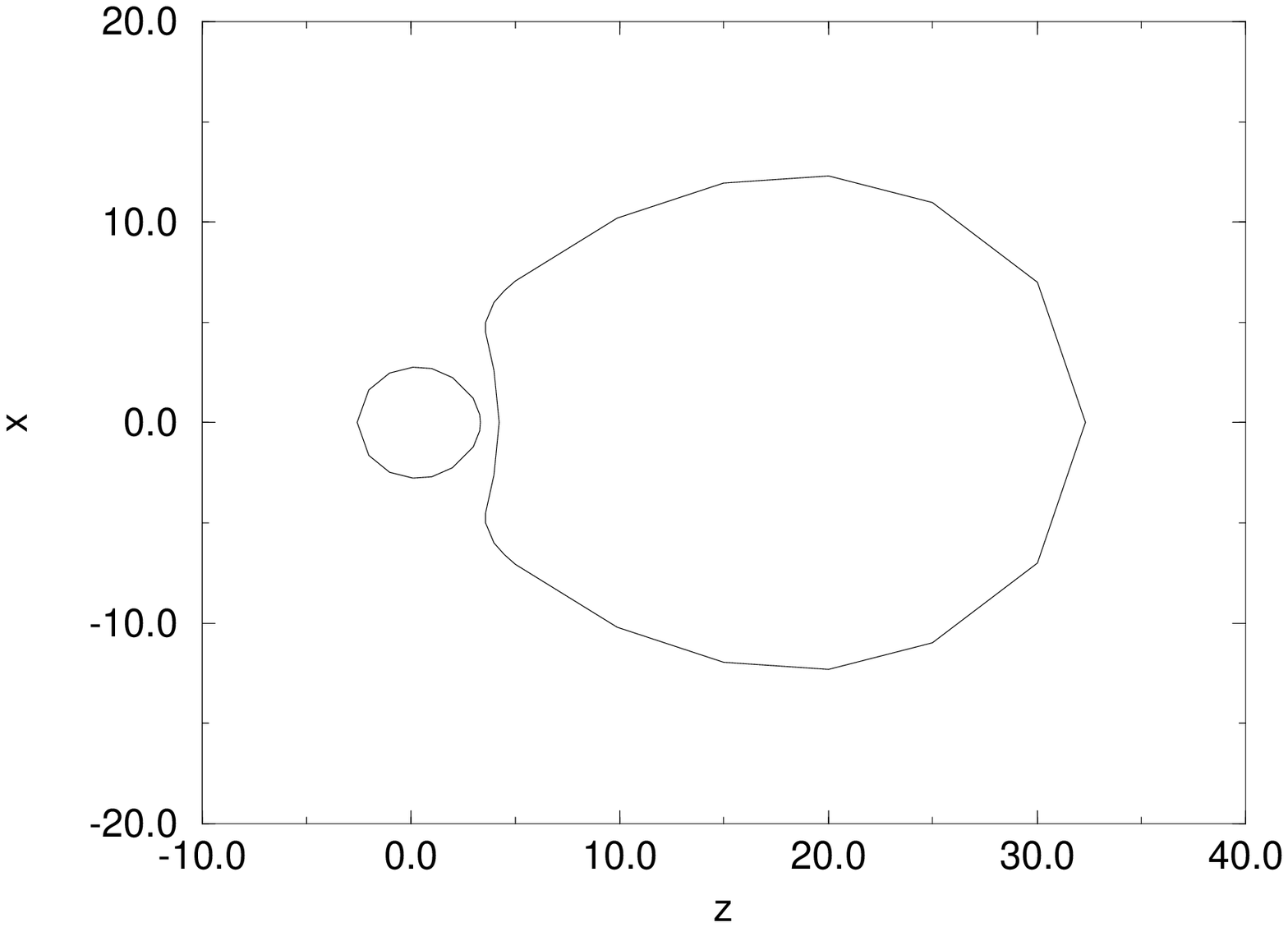,width=3in,height=2.5in}

\hspace{1.5in}\psfig{file=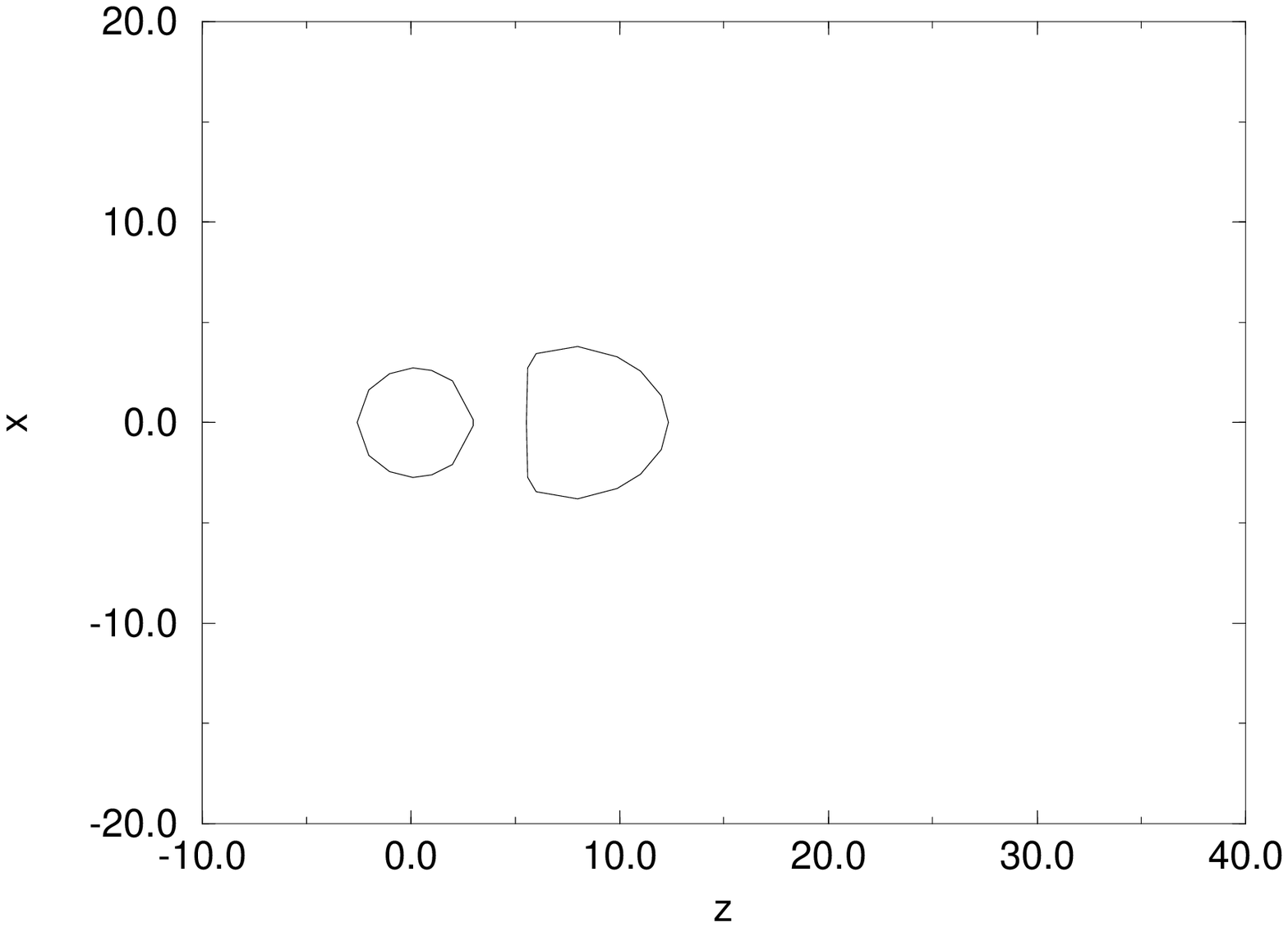,width=3in,height=2.5in}

\caption{Contour diagrams of the zero interaction energy density 
on the $x-z$ plane for $Sp(4)$ 
theory. $\alpha$ and $\beta$ monopoles are located at $(0, 0)$ and
$(0, 10)$, mass ratios $m_{\alpha}:m_{\beta}$ are choosen to be
(from top to bottom) 4, 19, 66, 99}
\end{figure}

From the contour diagrams one can see the major difference between 
the two theories when approaching the massless limit. In both cases we 
start with a simply connected contour for the small mass ratio  
(when the distance between two monopoles is large, the starting contour
could be different). The contour deforms and grows when the 
mass ratio increases, in both cases 
it breaks into two disjoint pieces when mass ratio 
is sufficiently large. The reason it breaks can be understood by
directly analyzing the massless limit of $\inter$ (in the $SU(3)$ case
$\inter$ itself vanishes but one can use $\inter /m_{\beta}$
which remains finite), Figs 7, 8 show those limits:
\begin{figure}[h]
\hspace{1.5in}\psfig{file=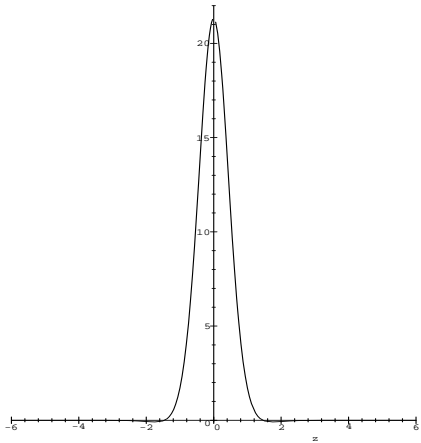,width=3in,height=3in}
\caption{$\rho_{\rm int}/m_{\beta}$ in $SU(3)$ theory at the massless limit
(this curve is independent of $\dd$)}
\end{figure}
\begin{figure}[h]
\hspace{1.5in}\psfig{file=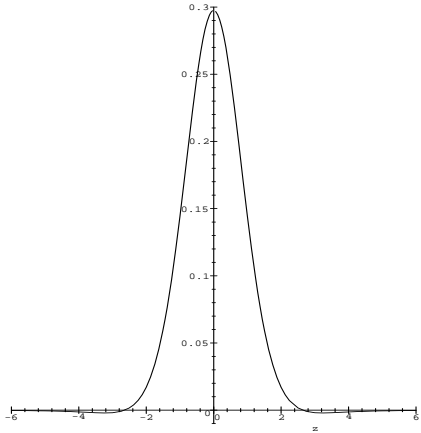,width=3in,height=3in}
\caption{$\rho_{\rm int}$ in $Sp(4)$ theory at the massless limit for $\dd=10$
(this curve has a weak dependence on $\dd$)}
\end{figure}
\noindent In both cases the limiting interaction
energy density becomes negative outside the core of the 
$\alpha$-monopole, this means that the contour can't keep growing 
and remains simply connected all the way as one increases the mass ratio.
After the breaking point, the part of the contour that surrounds the 
$\alpha$-monopole is stabilized, but the other part undergoes a 
very different evolution in the two cases: In the $SU(3)$ case that 
part of the contour
keeps growing, but gradually moves away from the $\alpha$-monopole 
(because of the scale of the graphs, it might not 
be able to see that easily, but
the shortest distance between the two parts of the contour is increasing). 
In the $Sp(4)$ case, however, the situation is the opposite, the other
part of the contour eventually shrinks and finally disappears 
(this happens before going to the limit). It should be mentioned that 
in $Sp(4)$ case if the $\beta$-monopole is inside the core of the 
$\alpha$-monopole, the contour could shrink and be stabilized without
breaking into two pieces at first.

The difference in the contour diagram of the two cases 
explains their different massless limits and accounts for 
the formation of the non-Abelian cloud.
Although in both cases, the interaction alters the energy distribution by
accumulating energy in certain regions, in the case of $SU(3)$ that
region is ever expanding and the effect of the interaction 
(therefore the massless monopole itself) is smeared out
over the infinitely large area, so the final energy density is completely 
dominated by the remaining massive monopole. On the other hand, 
in the $Sp(4)$ case, the interaction extracts energy and deposits it into a 
small region (in some sense one can say that 
the interaction is more ``localized'' in this case), 
as a result it affects the energy density distribution  
significantly, and because of the interaction the massless monopole doesn't 
grow into infinite size as it would if isolated. The 
non-Abelian cloud is just the effect of such an interaction.

\section{Conclusions}

So far we have analyzed the energy density of two monopole systems in
$SU(3)$ and $Sp(4)$ theories and obtained some idea on how the 
massless cloud forms. Based on these results one can make 
some qualitative conjectures on what might happen in general cases
where the interaction energy can be defined as 
\beq
\rho_{{\rm int}}=\rho_{{\rm total}}-\rho_{{\rm massive}}-
\rho_{{\rm massless}}
\eeq{52}
The basic property of $\inter$ we learned from the previous observation   
is that when the total magnetic charge is purely Abelian, the interaction
is more ``localized'' in contrast to the opposite case. Such an interaction 
extracts energy from distant regions, accumulates it in the 
vicinity of massive monopoles and gradually 
builds up the structure of the non-Abelian
cloud. This is fairly similar to the $Sp(4)$ case.

When the total magnetic charge is non-Abelian, however,  
qualitatively different situations could arise in general. To see this, 
let's look at the case with the symmetry breaking pattern 
$SU(N)\rightarrow U(1)\times SU(N-2)\times U(1)$ ($N>4$). Let 
$\alpha_1, ..., \alpha_{N-1}$ denote simple roots. When the system 
contains massive $\alpha_1$, $\alpha_{N-1}$ and would be massless 
$\alpha_i$ ($i=2, ..., N-3$) monopoles (notice that the 
$\alpha_{N-2}$-monopole 
is absent and so the total magnetic charge of the system is non-Abelian), 
only massive monopoles survive at the massless limit. This can be seen by 
noticing that the system under consideration is equivalent to the system 
studied in \cite{erick3} (which contains the $\alpha_{N-2}$-monopole 
as well and so the total
magnetic charge is Abelian) with the $\alpha_{N-2}$-monopole
removed. From \cite{erick3} 
we know that the only cloud parameter of the system is given by 
\beq
\dd=\sum_{i=2}^{N-1} |{\bf x}_i-{\bf x}_{i-1}| 
\eeq{99}
So removing any massless monopole is equivalent  
to removing the whole cloud (since it
makes the cloud size infinity) therefore only massive monopoles survive. This
situation is similar to the $SU(3)$ case we have studied. But there are
other systems which don't show such a direct analogue. As an example we can 
go back to our $Sp(4)$ theory, and consider a system with $N$ 
massive $\alpha$-monopoles and $N-1$ would be massless $\beta$-monopoles
(so the total magnetic charge is non-Abelian). At massless limit, the 
massless monopoles in this system will form a non-Abelian cloud rather
than disappearing. This is because such system 
can be obtained by removing
one $\beta$-monopole from a system containing $N$ $(N>1)$ 
$\alpha -\beta$ pairs. At the massless limit the latter system 
contains a non-Abelian cloud with many independent
size parameters. Removing one $\beta$-monopole will not make
all these parameters infinity and therefore will not destroy the whole cloud. 
Another way to understand this is to notice that when the extra 
$\alpha$-monopole is removed, we are left with a system made of 
$N-1$ $\alpha -\beta$ pairs (it can be called an Abelian sub-system)
which certainly contains the non-Abelian cloud. Since 
removing the $\alpha$-monopole won't create cloud, the cloud must exist
in the original system. This argument can be generalized.

These examples reveal the complexity of the general cases. It seems that
at the massless limit a system with a non-Abelian total magnetic charge 
can still contain massless monopoles 
(in the form of a non-Abelian cloud) in a ``maximal
Abelian sub-system''. We think further considerations on such 
situations will be interesting.

\vspace{5mm}

{\it Note added}: While writing this paper, we 
noticed the appearance of \cite{baal1}
from which the energy density of $SU(3)$ and $Sp(4)$ 
monopoles can also be obtained.

\vspace{5mm}

\centerline{\bf Acknowledgments}

The author wishes to thank Kimyeong Lee for useful discussions. 
This work is supported in part by the U.S. Department of Energy.

\end{document}